\documentstyle[12pt,frascatiphys]{article}

\newcommand{\kk}{\mbox{$K^+K^-$ }}
\newcommand{\ksks}{\mbox{$K^0_SK^0_S$ }}

\newcommand{\pipi}{\mbox{$\pi^+\pi^-$ }}

\begin{document}
\title{
NEW RESULTS FROM EXPERIMENT WA102: \\
A PARTIAL WAVE ANALYSIS
OF THE CENTRALLY PRODUCED $K \overline K $ AND $\pi \pi$ SYSTEMS
}
\author{
A. Kirk        \\
{\em School of Physics and Astronomy, University of Birmingham, U.K.} \\
}
\maketitle
\baselineskip=14.5pt
\begin{abstract}
A partial wave analysis of the centrally produced $K \overline K$ and $\pi \pi$
systems shows that the $f_J(1710)$ has J~=~0.
In addition,
a study of central meson production as a function of the
difference in transverse momentum ($dP_T$)
of the exchanged particles
shows that undisputed $q \overline q$ mesons
are suppressed at small $dP_T$ whereas the glueball candidates
are enhanced and that
the production cross section for different
resonances depends strongly on the azimuthal angle between the
two outgoing protons.
\end{abstract}
\baselineskip=17pt
\section{Introduction}
\par
There is considerable current interest in trying to isolate the lightest
glueball.
Several experiments have been performed using glue-rich
production mechanisms.
One such mechanism is Double Pomeron Exchange (DPE) where the Pomeron
is thought to be a multi-gluonic object.
Consequently it has been
anticipated that production of
glueballs may be especially favoured in this process\cite{closerev}.
\par
The WA102 experiment at the CERN Omega Spectrometer
studies centrally produced exclusive final states
formed in the reaction
\noindent
\begin{equation}
pp \longrightarrow p_{f} X^{0} p_s,
\label{eq:1}
\end{equation}
where the subscripts $f$ and $s$ refer to the fastest and slowest
particles in the laboratory frame respectively and $X^0$ represents
the central system.
\section{A partial wave analysis of the $K \overline K$ system}
\par
\begin{figure}[h]
 \vspace{7.0cm}
\begin{center}
\includegraphics{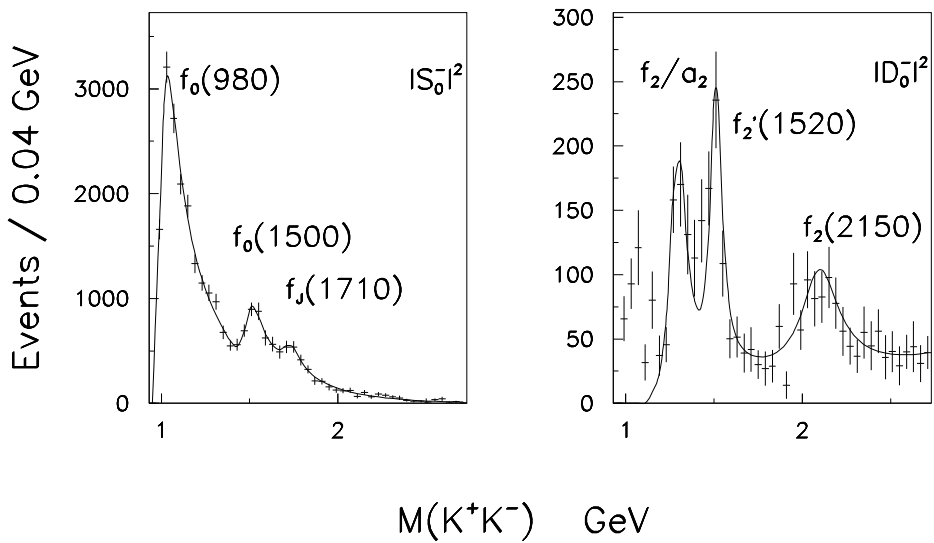}
\end{center}
\caption{\it The $S_0^-$ and $D_0^-$-Waves
resulting from a partial wave analysis of
the $K^+K^-$ system.}
\label{fi:1}
\end{figure}

The isolation of the reaction
\begin{equation}
pp \rightarrow p_{f} (K^+ K^-) p_{s}
\label{eq:b}
\end{equation}
has been described in detail in a previous publication\cite{re:kkpap}.
A Partial Wave Analysis (PWA) of the centrally produced \kk system has been
performed,
using the reflectivity basis\cite{chung},
in 40~MeV intervals of the \kk
mass spectrum using an event-by-event maximum likelihood
method\cite{re:kkpap}.
The $S_0^-$ and $D_0^-$-Waves
from the physical solution are shown in fig.~\ref{fi:1}.
\par
The $S_0^-$-wave shows a threshold enhancement; the peaks at 1.5 GeV and
1.7~GeV are interpreted as being due to the
$f_0(1500)$ and $f_J(1710)$ with J~=~0.
A fit has been performed to the $S_0^-$ wave using
three interfering Breit-Wigners to describe the $f_0(980)$, $f_0(1500)$
and $f_J(1710)$ and a background
of the form
$a(m-m_{th})^{b}exp(-cm-dm^{2})$, where
$m$ is the
\kk
mass,
$m_{th}$ is the
\kk
threshold mass and
a, b, c, d are fit parameters.
The resulting fit is shown in fig.~\ref{fi:1} and gives
for the $f_0(980)$ M~=~985~$\pm$~10~MeV, $\Gamma$~=~65~$\pm$~20~MeV,
for the $f_0(1500)$ M~=~1497~$\pm$~10~MeV, $\Gamma$~=~104~$\pm$~25~MeV and
for the $f_0(1710)$ M~=~1730~$\pm$~15~MeV, $\Gamma$~=~100~$\pm$~25~MeV
parameters which are consistent with the PDG\cite{PDG98} values for these
resonances.
\par
The $D_0^-$-wave shows peaks in the 1.3 and 1.5~GeV regions,
presumably due to the $f_2(1270)/a_2(1320)$ and $f_2^\prime(1525)$ and
a wide structure above 2 GeV. There is no evidence for
any significant structure in the D-wave in the region of the
$f_J(1710)$. In addition, there are no statistically significant
structures in any of the other waves.
A fit has been performed to the $D_0^-$ wave above 1.2~GeV using
three incoherent relativistic spin 2 Breit-Wigners
to describe the $f_2(1270)/a_2(1320)$,
$f_2^\prime(1525)$ and the peak at 2.2 GeV and
a background of the form described above.
The resulting fit is shown in fig.~\ref{fi:1} and gives
for the $f_2(1270)/a_2(1320)$ M~=~1305~$\pm$~20~MeV,
$\Gamma$~=~132~$\pm$~25~MeV,
for the $f_2^\prime(1525)$ M~=~1515~$\pm$~15~MeV, $\Gamma$~=~70~$\pm$~25~MeV
and for the
$f_2(2150)$ M~=~2130~$\pm$~35~MeV, $\Gamma$~=~270~$\pm$~50~MeV.
\par
A study has also been made of the centrally produced \ksks
channel\cite{re:kkpap}.
This channel has lower statistics than the \kk channel but has
the advantage that only even spins can contribute, which also means that
there are only two ambiguous solutions to the PWA.
The physical solution resulting from the PWA is
the same as for the \kk final state; namely
that
the $S_0^-$-wave shows a threshold enhancement and peaks at 1.5 GeV and
1.7~GeV interpreted as being due to the
$f_0(1500)$ and $f_J(1710)$ with J~=~0.
\section{A partial wave analysis of the $\pi \pi$ system}
\begin{figure}[h]
 \vspace{9.0cm}
\begin{center}
\includegraphics{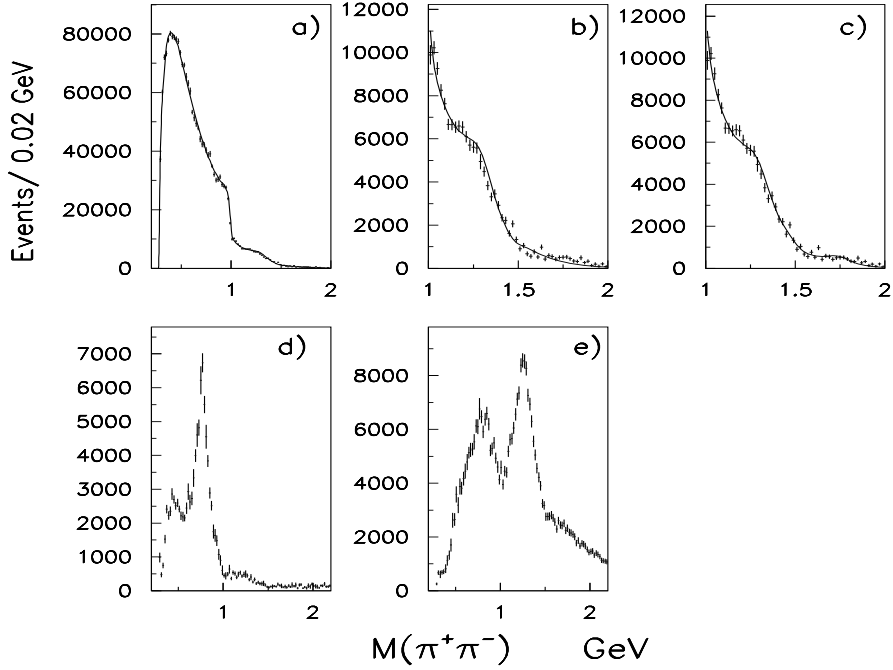}
\end{center}
\caption{\it a), b), c) The $S_0^-$ wave d) the $P_0^-$ wave and
e) the $D_0^-$ wave
resulting from a partial wave analysis of
the $\pi^+\pi^-$ system.}
\label{fi:2}
\end{figure}
\par
The isolation of the reaction
\begin{equation}
pp \rightarrow p_{f} (\pi^+ \pi^-) p_{s}
\label{eq:c}
\end{equation}
has been described in detail in a previous publication\cite{re:pipipap}.
The resulting centrally produced \pipi system
consists of 2.87 million events.
A PWA of the centrally produced \pipi system has been
performed,
using the reflectivity basis\cite{chung},
in 20~MeV intervals of the \pipi
mass spectrum using an event-by-event maximum likelihood
method\cite{re:pipipap}.
The $S_0^-$, $P_0^-$ and $D_0^-$-Waves
from the physical solution are shown in fig.~\ref{fi:2}.
The $S_0^-$-wave spectrum
shows a clear threshold enhancement followed by a sharp
drop at 1~GeV. There is clear evidence for the $\rho(770)$ in the $P_0^-$ wave
and for the $f_2(1270)$ in the
$D_0^-$ wave.
\par
An interesting feature of the $D_0^-$ wave is the presence of a structure
below 1~GeV.
In order to see if this effect is due to acceptance problems or
problems due to non-central events, we have reanalysed the data
using a series of different cuts
but after acceptance
correction no cut has been found that can remove
the low mass structure.
In order to investigate any systematic effects we have also analysed the
central $\pi^0\pi^0$ data and a similar structure is also
found\cite{re:pi0pi0pap}.
This structure does indeed seem to be a real effect
which is present in other centrally produced $\pi \pi$ systems\cite{cenprod}.
\par
In order to obtain a satisfactory fit
to the $S_0^-$ wave from threshold to 2~GeV it has been found to be
necessary to use
three interfering Breit-Wigners to describe the $f_0(980)$, $f_0(1300)$
and $f_0(1500)$ and a background
of the form
$a(m-m_{th})^{b}exp(-cm-dm^{2})$, where
$m$ is the
\pipi
mass,
$m_{th}$ is the
\pipi
threshold mass and
a, b, c, d are fit parameters.
The fit is shown in fig.~\ref{fi:2}a) for the entire mass range
and in fig.~\ref{fi:2}b) for masses above 1 GeV.
The resulting parameters are
for the $f_0(980)$ M~=~982~$\pm$~3~MeV, $\Gamma$~=~80~$\pm$~10~MeV, for the
$f_0(1300)$ M~=~1308~$\pm$~10~MeV, $\Gamma$~=~222~$\pm$~20~MeV and for the
$f_0(1500)$ M~=~1502~$\pm$~10~MeV, $\Gamma$~=~131~$\pm$~15~MeV
which are consistent with the PDG~\cite{PDG98} values for these
resonances.
As can be seen, the fit describes the data well for masses below 1~GeV.
It was not possible to describe the data above 1~GeV without the addition
of both the $f_0(1300)$ and $f_0(1500)$ resonances.
However, even with this fit using
three Breit-Wigners it can be seen that the fit does not
describe well the 1.7 GeV region.
This could be due to a \pipi decay mode of the $f_J(1710)$ with J~=~0.
Including a fourth Breit-Wigner in this mass region decreases the
$\chi^2$ from 256 to 203 and yields
for the $f_J(1710)$ M~=~1750~$\pm$~20~MeV and $\Gamma$~=~160~$\pm$~30~MeV
parameters which are consistent with the PDG~\cite{PDG98} values for the
$f_J(1710)$.
The fit is
shown in fig.~\ref{fi:2}c) for masses above 1 GeV.
\section{A Glueball-$q \overline q$ filter in central production ?}
The WA102 experiment studies mesons produced in double exchange processes.
However, even in the case of pure DPE
the exchanged particles still have to couple to a final state meson.
The coupling of the two exchanged particles can either be by gluon exchange
or quark exchange. Assuming the Pomeron
is a colour singlet gluonic system if
a gluon is exchanged then a gluonic state is produced, whereas if a
quark is exchanged then a $q \overline q $ state is produced\cite{closeak}.
In order to describe the data in terms of a physical model,
Close and Kirk\cite{closeak},
have proposed that the data be analysed
in terms of the difference in transverse momentum ($dP_T$)
between the particles exchanged from the
fast and slow vertices.
The idea being that
for small differences in transverse momentum between the two
exchanged particles
an enhancement in the production of glueballs
relative to $q \overline q$ states may occur.
\begin{figure}
 \vspace{7.0cm}
\begin{center}
\includegraphics{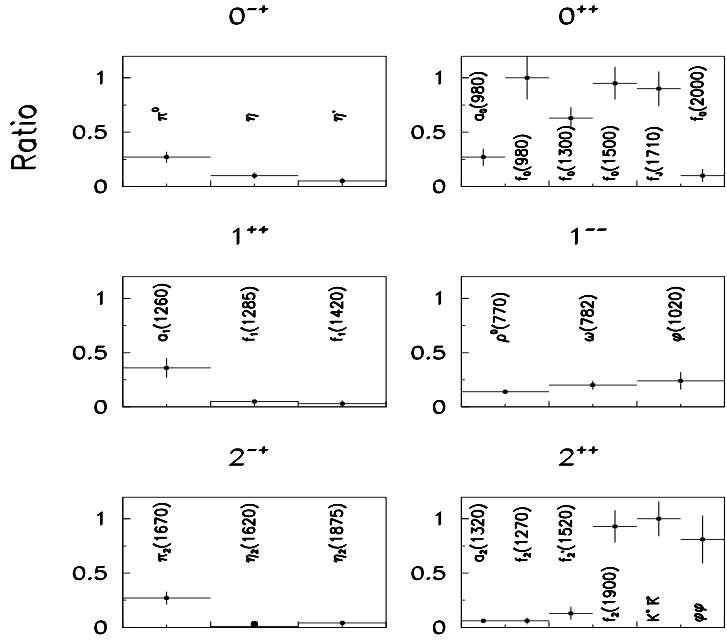}
\end{center}
\caption{\it The ratio of the amount of resonance with
$dP_T$~$\leq$~0.2 to the amount with
$dP_T$~$\geq$~0.5~GeV.
}
\label{fracratio}
\end{figure}
\par
The contribution of each resonance as a function
of $dP_T$ has been calculated.
Figure~\ref{fracratio} shows the ratio of the number of events
for $dP_T$ $<$ 0.2 GeV to
the number of events
for $dP_T$ $>$ 0.5 GeV for each resonance considered.
It can be observed that all the undisputed $q \overline q$ states
which can be produced in DPE, namely those with positive G parity and $I=0$,
have a very small value for this ratio ($\leq 0.1$).
Some of the states with $I=1$ or G parity negative,
which can not be produced by DPE,
have a slightly higher value ($\approx 0.25$).
However, all of these states are suppressed relative to the
the glueball candidates the
$f_0(1500)$, $f_J(1710)$, and $f_2(1930)$,
together with the enigmatic $f_0(980)$,
which have
a large value for this ratio.
\begin{figure}
 \vspace{7.0cm}
\begin{center}
\includegraphics{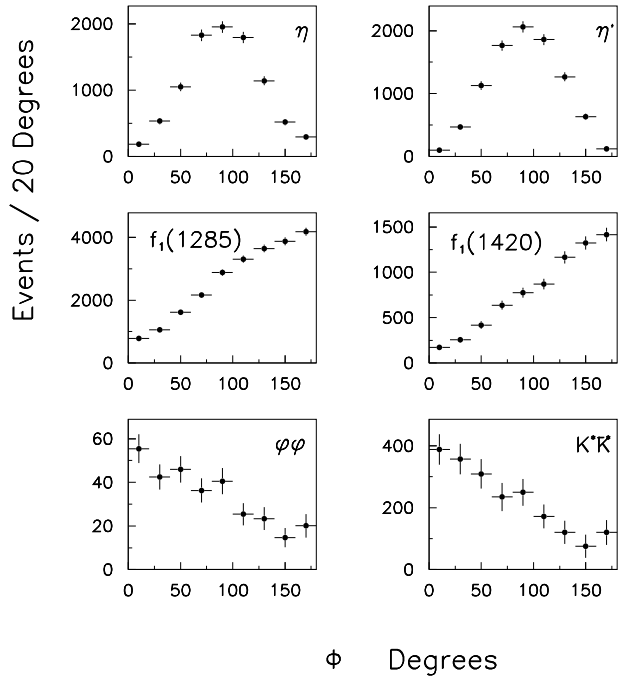}
\end{center}
\caption{\it The azimuthal angle between the fast and
slow protons ($\phi$) for various final states.
}
\label{fi:phidep}
\end{figure}

\section{The azimuthal angle between the outgoing protons}

\par
The azimuthal angle ($\phi$) is defined as the angle between the $p_T$
vectors of the two protons.
Naively it may be expected that this angle would be flat irrespective
of the resonances produced.
Fig.~\ref{fi:phidep}  shows the $\phi$ dependence for two
$J^{PC}$~=~$0^{-+}$ final states (the $\eta$ and $\eta^\prime$),
two $J^{PC}$~=~$1^{++}$ final states (the $f_1(1285)$ and $f_1(1420)$) and
two $J^{PC}$~=~$2^{++}$ final states
(the $\phi \phi$ and $K^*(892) \overline K^*(892)$ systems).
The $\phi$ dependence is clearly not flat and considerable variation
is observed between final states with different $J^{PC}$s.

\section{Summary}
\par
In conclusion, a partial wave analysis of the centrally
produced $K \overline K$ system has been performed.
The striking feature is the
observation of peaks in the $S_0^-$-wave corresponding to
the $f_0(1500)$ and $f_J(1710)$ with J~=~0.
In addition, a partial wave analysis of a
high statistics sample of centrally produced \pipi events
shows that the $S_0^-$-wave is composed of
a broad enhancement at threshold, a sharp drop
at 1 GeV due to the interference between the $f_0(980)$
and the S-wave background, the $f_0(1300)$, the $f_0(1500)$ and
the $f_J(1710)$ with J~=~0.
\par
A study of centrally produced pp interactions
show that there is the possibility of a
glueball-$q \overline q$ filter mechanism ($dP_T$).
All the
undisputed $q \overline q $ states are observed to be suppressed
at small $dP_T$, but the glueball candidates
$f_0(1500)$, $f_J(1710)$, and $f_2(1930)$ ,
together with the enigmatic $f_0(980)$,
survive.
In addition, the production cross section for different
resonances depends strongly on the azimuthal angle between the
two outgoing protons.


\begin{thebibliography}{99}
\bibitem{closerev} D.Robson, Nucl. Phys. {\bf B130}, 328 (1977) ; \\
F.E. Close, Rep. Prog. Phys. {\bf 51}, 833  (1988) .
\bibitem{re:kkpap}
D. Barberis {\em et al.,} hep-ex/9903042.
\bibitem{chung}
S.U. Chung, Phys. Rev. {\bf D56}, 7299 \rm (1997).
\bibitem{PDG98}
Particle Data Group, European Physical Journal {\bf C3}, 1 (1998).
\bibitem{re:pipipap}
D. Barberis {\em et al.,} hep-ex/9903043.
\bibitem{re:pi0pi0pap}
D. Barberis {\em et al.,} hep-ex/9903044.
\bibitem{cenprod}
D. Alde {\em et al.,} (NA12/2 collaboration) In preparation; \\
M.C. Berrisso {\em et al.,} (E690 Collaboration) Proceedings of Hadron 97, AIP
conf. proc. 432 (1997) 36; \\
T. $\AA$kesson {\em et al.,} (AFS collaboration) Nucl. Phys. {\bf B264}, 154
(1986).
\bibitem{closeak}
F.E. Close and A. Kirk, Phys. Lett. {\bf B397}, 333 (1997).

\end{thebibliography}
\end{document}